\documentstyle[12pt,epsf]{article}
\def\gsim{\mathrel{\rlap {\raise.5ex\hbox{$ > $}}
{\lower.5ex\hbox{$\sim$}}}}
\def\lsim{\mathrel{\rlap {\raise.5ex\hbox{$ < $}}
{\lower.5ex\hbox{$\sim$}}}}

\newcommand{\be}{\begin{equation}}
\newcommand{\ee}{\end{equation}}
\newcommand{\bea}{\begin{eqnarray}}
\newcommand{\nn}{\nonumber}
\newcommand{\eea}{\end{eqnarray}}

\def\gappeq{\mathrel{\rlap {\raise.5ex\hbox{$>$}}
{\lower.5ex\hbox{$\sim$}}}}
 
\def\lappeq{\mathrel{\rlap{\raise.5ex\hbox{$<$}}
{\lower.5ex\hbox{$\sim$}}}}
 
\begin{document}
 
\begin{titlepage}
\begin{flushright}
ACT-9/98\\
CTP-TAMU-39/98 \\
OUTP--98--74P \\
gr-qc/9810086 \\
\end{flushright}

\begin{centering}
\vspace{.1in}

{\large {\bf Time-Dependent
Vacuum Energy Induced by $D$-Particle Recoil }} \\ 
\vspace{.2in}

{\bf John Ellis}$^{a}$, {\bf N.E. Mavromatos}$^{b}$ 
and {\bf D.V. Nanopoulos}$^{c,d,e}$ \\

\vspace{.5in}
 
{\bf Abstract} \\
\vspace{.1in}
\end{centering}
{\small We consider cosmology in the framework of a `material reference
system' of $D$ particles, including the effects of quantum recoil
induced by closed-string probe particles. We find a
time-dependent contribution to the cosmological 
vacuum energy, which relaxes to zero as $\sim 1/ t^2$ for large
times $t$. If this energy density is dominant, 
the Universe expands with a scale factor $R(t) 
\sim t^2$. We show that this possibility is
compatible with recent observational constraints from
high-redshift supernovae, and may also respect other phenomenological
bounds on time variation in the vacuum energy imposed by
early cosmology. }
\vspace{1.in}
\begin{flushleft}
$^{a}$ Theory Division, CERN, CH 1211 Geneva 23, Switzerland. \\
$^{b}$ Department of Physics, Theoretical Physics,
University of Oxford, 1 Keble Road, 
Oxford OX1 3NP, U.K.  \\
$^{c}$ Center for
Theoretical Physics, Dept. of Physics,
Texas A \& M University, College Station, TX 77843-4242, USA, \\
$^{d}$ Astroparticle Physics Group, Houston
Advanced Research Center (HARC), The Mitchell Campus,
Woodlands, TX 77381, USA. \\
$^{e}$ Academy of Athens, Chair of Theoretical Physics, 
Division of Natural Sciences, 28 Panepistimiou Avenue, 
Athens 10679, Greece. \\

\end{flushleft}

\end{titlepage}

\newpage

\section{Introduction}

The possibility that the vacuum, the lowest-energy state, 
might not actually have zero energy was first raised by
Einstein~\cite{einstein}, 
who regarded this proposal as his greatest mistake.
Various possible contributions to this vacuum energy are known
in field theory, including contributions associated with
condensates in QCD and the electroweak theory that are many
orders of magnitude larger than the possible physical value
of the cosmological constant today, and quantum contributions 
that are formally highly divergent. The existences of these
possible contributions to the vacuum energy mean that the
issue of a possible cosmological constant cannot 
be avoided~\cite{turner,cosmo},
although this possibility can only be addressed theoretically in a
complete quantum theory of gravity.

There have been attempts to address the issue of the
cosmological constant in various approaches to quantum
gravity~\cite{cosmo}. The cogency of many of these approaches was
limited by the presence of unrenormalizable quantum divergences,
but some interesting ideas emerged, including the suggestion that
the Universe might be relaxing towards a zero-energy state during
the course of its cosmological expansion~\cite{relaxation}. 
String theory is
the only known candidate for a completely consistent
theoretical framework for quantum gravity, and has offered
several new insights into the cosmological-constant problem.
For example, the cosmological constant vanishes in a
supersymmetric string theory, ideas have been proposed for
concealing supersymmetry in the observable world~\cite{nosusy}, 
and suggestions
have been made how the cosmological constant might vanish
even in the absence of supersymmetry~\cite{nosusystr}.

The issue of possible vacuum energy has been cast in a new light
by recent astrophysical observations suggesting that it might 
indeed be non-zero. The theory of cosmological inflation
strongly suggests that the current density of the Universe is
close to the critical value: $\Omega_{TOT} = 1$, and this is
supported by the location of the first acoustic peak,
whose existence is hinted at by data on fluctuations in the
cosmic microwave background radiation~\cite{COBE}. On the other hand, the
matter density inferred from data on large-scale structures~\cite{BF} in
the
Universe does not rise above $\Omega_{M} \sim 0.3$. This
includes the baryonic density, which is believed on the basis of
cosmological nucleosynthesis arguments to be much smaller:
$\Omega_{B} < 0.1$. Most of the matter density is thought to consist
of cold dark matter, but this is not sufficient by itself to explain
all the data on microwave background fluctuations and large-scale
structure formation~\cite{GS}. Hot dark matter was for some time
a favoured epicycle for cold dark matter, but this would also need to be
included within the $\Omega_{M} \sim 0.3$ inferred from observations
of large-scale structure. Moreover, the recent data on atmospheric 
and solar neutrino oscillations do not suggest a neutrino mass 
large enough to contribute significantly to $\Omega_{M}$~\cite{SK}. Taken
together, these arguments reopen the possibility of a cosmological
constant $\Lambda$: $\Omega_{\Lambda} \sim 0.7$.

This possibility has recently received dramatic support from an
unexpected source, namely observations of high-redshift supernovae~\cite{sn}.
These indicate that the large-scale geometry of the Universe is not
that of a critical matter-dominated cosmology, and that its
expansion may even be accelerating. The supernova data are consistent
with $\Omega_{\Lambda} \sim 0.7$, if the Universe is indeed close to
critical as suggested by inflation. It should be stressed that these
observations are entirely independent of the earlier arguments 
given in the previous paragraph, encouraging us to take seriously the
possibility that the vacuum energy density may be non-zero.

This presents theoretical physics with a tremendous opportunity:
a number to be calculated within one's quantum theory of gravity,
that can be confronted with measurement. Many of the previous
discussions of the cosmological constant had included attempts
to show why it vanishes. Maybe it does not? The known exact
symmetries are not adequate to derive $\Lambda = 0$, and it may be
small because of some approximate symmetry: for example, the
value allowed by the supernova data corresponds to $\Lambda \sim
(M_W / M_P )^8$ in natural units. Alternatively, perhaps the 
vacuum energy is relaxing towards zero~\cite{relaxation}: for example, the
present
age of the Universe $t_0 \sim 10^{60}$ in natural units, so
perhaps the vacuum energy is decreasing as $\Lambda \sim (1/t_0)^2$?
We present just such a scenario in this paper.

Our starting-point is the expectation that the vacuum contains
Planck-scale quantum fluctuations in topology, on the Planck
time scale and with Planckian density. Working in the context
of string theory, in which this space-time foam may be
described~\cite{dbrecoil,kanti} 
using $D$-brane technology~\cite{dbranes}, we are not in a position to
calculate absolutely the limiting value of the vacuum energy density.
However, we are able to isolate a contribution to the vacuum energy
that decreases like $(1/t)^2$, providing a mechanism for relaxation
towards a limiting value that may well vanish. This contribution is
due to the quantum recoil of $D$ branes in the space-time foam,
which exhibit energy excitations that are interpreted classically
as a time-dependent energy density. We show that the time-dependence
we find is compatible with the constraints imposed by the high-redshift
supernova data~\cite{sn}, as well as with the values of $\Lambda$ 
allowed earlier in the history of the Universe.

\section{Material Reference Frame of $D$ Particles, Recoil and
Anti-de-Sitter Space}

The use of a material reference system (MRS) in General Relativity 
has a long history. First conceived 
by Einstein~\cite{einstein} 
and Hilbert~\cite{hilbert} in the form of a system of rods and clocks,
MRS have been subsequently used as a general tool to 
specify events in space time and to address conceptual questions in 
General Relativity and later in Quantum Gravity~\cite{dewitt},
particularly
in connection with the implications of the uncertainty principle 
for measurements of the gravitational field. 
In this latter respect, we mention 
arguments~\cite{rovelli}
that the quantum properties of the bodies that form a MRS are responsible 
for making physical operators in Quantum Gravity well defined. 

A useful example of a MRS is that of a relativistic 
elastic medium considered by DeWitt~\cite{dewitt}. 
Its action is :
\be
  S_{mrs} [x^a;g_{ab}] = \int d\sigma \int _{S^{(3)}} d^3\zeta 
\{ -\left(n M + w \right)\sqrt{-{\dot x}^a {\dot x}^bg_{ab} (x)}\} 
\label{mrs}
\ee
where $S^{(3)}$ is the `matter' spatial manifold, whose points $\zeta \in
S^{(3)}$ label the particle world lines, the variables  
$x^a$ denote the coordinates 
of a relativistic particle probe of mass $M$
moving in the MRS, which, together with 
the background metric $g_{ab}$, are considered 
functions of $\sigma$ and $\zeta ^i$. 
The quantity $n$ denotes the particle-number density, 
whilst $w$ is the interaction-energy density in the 
comoving frame. The system described by (\ref{mrs}) 
is reparametrization invariant,
i.e., it is invariant under the infinitesimal transformations $\delta x^a
= -\epsilon {\dot x}^a $ induced by reparametrizations $\sigma \rightarrow
\epsilon (\sigma, \zeta )$ of the particle world lines.

The above example is a prototype for our case, where 
we consider
an ensemble of Dirichlet $D$ branes~\cite{dbranes} as
a MRS through which 
closed-string matter propagates. 
We assume the existence of a suitable conformal 
closed-string theory in $D=10$ or $11$ 
dimensions~\footnote{An eleven-dimensional manifold arises 
naturally when one incorporates world-sheet defects~\cite{mth}. For our
purposes in this paper, the initial dimension of the string
theory is not
relevant, as long as it is at least ten.}
that admits $D$-brane solutions. 
These solitonic objects are located at
fixed points in target space, and hence are suitable for
defining a MRS.

We now consider a configuration combining a closed-string state (matter) 
and a $D$ particle, which induces a recoil distortion of the
$D$ brane describable within a
conformal field theory setting as in~\cite{kmw}. 
The recoil is best described by 
the splitting of the closed-string matter state into two
open-string states with their ends attached on the $D$ brane. 
In the world-sheet formalism, the recoil is described~\cite{kmw} 
by a suitable pair of logarithmic operators~\cite{gurarie}, 
corresponding to the collective coordinates $y^i$ and velocities $u_i$ 
of the recoiling $D$ particles.  
Such a scattering procedure constitutes a generalization of the Heisenberg
microscope approach, where the r\^ole of Heisenberg's photon is played
by the closed-string state, whilst the system of $D$ branes
plays the r\^ole of the detector (or measuring apparatus). 

As already mentioned, we concentrate on a single scattering event, namely
the scattering of a single closed-string state by a single defect. 
We are unable at present to treat fully the
more realistic
case of an ensemble of defects with Planckian density, 
due to our limited understanding of the underlying 
microscopic dynamics. Instead, we interpret the 
single-scattering approach as the first step
in a dilute-gas approximation for the $D$ particles, which should
be sufficient to describe qualitatively the 
leading behaviour of the vacuum energy of the Universe. 

The combined system is
characterized by a homotopic `evolution' parameter ${\cal T}$.
We look for a consistent description of the
coupled system in a maximally-symmetric background space,
which includes the pair of logarithmic 
deformations that
correspond to the $D$-dimensional location $y_i$
of the recoiling $D$ brane and the
homotopic `velocity' $u_i \equiv \partial_{\cal T}
y_i$~\cite{kmw,lizzi}.
These two operators are slightly relevant~\cite{kmw}, in a 
world-sheet renormalization-group sense, with anomalous dimensions
$\Delta = -{\epsilon ^2 / 2}$  
where $\epsilon \rightarrow 0^+$ is a regularization parameter.
This is independent of the homotopic `velocity' $u_i$, but is
related~\cite{kmw} to the world-sheet size $L$ 
and a world-sheet short-distance cut-off $a$ via 
\be
\epsilon ^{-2} \sim \eta {\rm ln}(L/a)^2,
\label{epsilon}
\ee
where $\eta = \pm 1$ 
for a Euclidean- (Minkowski-)signature homotopic parameter ${\cal T}$. 
Thus, the recoiling $D$ brane is no longer described by a 
conformal theory on the world sheet, despite the fact that
the theory was conformally invariant before the encounter
that induced the recoil.

To restore 
conformal invariance, we invoke 
Liouville dressing~\cite{ddk} by a
mode $\varphi$ that can be identified~\cite{dbrecoil,kanti,emn} 
with a {\it time-like} homotopic variable ${\cal T}$.
This Liouville field restores conformal
invariance in an initially critical string theory.
The dressing by such a time-like Liouville mode $\varphi \equiv
{\cal T}$ leads to an effective curved space-time manifold
in $D+1$ dimensions. We find a consistent solution
of the world-sheet $\sigma$-model equations of motion which is
described~\cite{kanti} by a metric of the form: 
\begin{equation}
G_{00}=-1 \,,\, G_{ij}=\delta_{ij} \,,\,
G_{0i}=G_{i0}=f_i(y_i, {\cal T})=\epsilon (\epsilon y_i + u_i {\cal T})\,
,\,\,i,j=1,...,D
\label{yiotametric}
\end{equation}
We restrict ourselves to the case  
where the recoil velocity $u_i \rightarrow 0$, as
occurs if the $D$ brane  
is very heavy. This is formally justified in
the weak-coupling limit for the string, since the
$D$-brane mass $M \propto 1/g_s$,
where $g_s \rightarrow 0$ is the string 
coupling. From the world-sheet point of view~\cite{emnmonop,mth}, 
such a very heavy $D$ brane corresponds 
to a strongly-coupled defect, since the coupling $g_v$ of the
world-sheet defect is related to the string coupling $g_s$ by
\be
     g_v \propto \frac{1}{\sqrt{g_s}} 
\label{duality}
\ee
This is a manifestation of
world-sheet/target-space strong/weak-coupling duality. 

In the limit $u_i \rightarrow 0$,  
the only non-vanishing components of the 
$D$-dimensional Ricci tensor are~\cite{kanti}: 
\be
R_{ii} 
\simeq 
\frac{-(D-1)/|\epsilon|^4}{(\frac{1}{|\epsilon|^4} - \sum_{k=1}^{D}|y_i|^2)^2} 
+{\cal O}(\epsilon ^{8}) 
\label{limRicci}
\ee
where we have taken (\ref{epsilon}) into account,
for the appropriate Minkowskian signature of the Liouville 
mode ${\cal T}$. In this limiting case, 
the Liouville mode decouples when ${\cal T} >>0$, 
and one is effectively left with a 
maximally-symmetric $D$-dimensional manifold. Hence, we
may write (\ref{limRicci}) as
\be 
   R_{ij}=\frac{1}{D}{\cal G}_{ij}R 
\label{newricci}
\ee
where ${\cal G}_{ij}$ is a diagonal metric corresponding to the line
element:
\be
      ds^2=\frac{|\epsilon|^{-8}\sum_{i=1}^{D} dy_i^2}{(\frac{1}{|\epsilon|^4} - \sum_{i=1}^{D}|y_i|^2)^2} 
\label{ball}
\ee
This metric describes the interior
of a $D$-dimensional ball, 
which is the Euclideanized version of an 
anti-de-Sitter (AdS) space time. In its Minkowski version,
one can easily check that the curvature corresponding to (\ref{ball}) is
\be
R = -4 D (D - 1) |\epsilon|^4,
\label{curvature}
\ee
which is {\it constant} and {\it negative}. The radius of the AdS
space is $b = |\epsilon|^{-2}$.

The Ricci tensor (\ref{limRicci}) corresponds
to the 
low-energy: ${\cal O}(\alpha ')$, $\alpha ' << 1$ equation of
motion for a world-sheet $\sigma$ model, 
as obtained from the
vanishing of the $\beta$ function in this background.
The Ricci tensor (\ref{limRicci})
cannot be a consistent string background
compatible with conformal invariance to order 
$\alpha '$ if only tree-level world-sheet 
topologies are taken into account. 
However, as shown in~\cite{fischler},
this conclusion no longer holds when
one includes string-loop corrections. These induce a target-space
cosmological constant, corresponding to a dilaton 
tadpole, which renders the backgrounds (\ref{limRicci}) consistent 
with the conformal-invariance conditions.

Alternatively, as discussed
in~\cite{kanti}, the cosmological vacuum energy
may be considered as being obtained from an effective
tree-level non-critical Liouville string 
with central-charge deficit 
\be 
     Q^2 = \Lambda \propto -2(\alpha  ')^2 
(D-1)(D-2)|\epsilon |^4 + {\cal O}(\epsilon ^6)
\label{ccd}
\ee
As we argue in the next section, this leads to a
non-trivial time-dependent vacuum energy when we identify
$\epsilon ^2$ with a temporal evolution variable, after appropriate 
analytic continuation to imaginary values. 
The analytic continuation restores positivity of the 
deficit $Q^2$, as is appropriate for supercritical string
models~\cite{aben}. 

\section{Interpretation as Physical Vacuum Energy}

In order to discuss the physical interpretation of the above analysis,
we consider the $D$-dimensional 
components $G_{ij}$ of the $\sigma$-model metric (\ref{ball})  
to be purely {\it spatial}. In that case we may identify the 
analytically continued $i\epsilon ^2 $ as a Liouville `physical' time 
$t$, 
\be 
     i\epsilon^{-2} \rightarrow t 
\label{physical}
\ee
By construction~\cite{ddk}, the resulting Universe is 
of Friedmann-Robertson-Walker (FRW) type, 
since the $\sigma$-model 
kinetic terms for the Liouville field $\phi$ are of the form
$\int d^2\sigma (-\partial \phi {\overline \partial}\phi ) $, 
corresponding to a time-like component of the metric
of the form:
\be
         G_{00}=-1
\label{temporal}
\ee
This Minkowskian signature is a consequence of
the fact that the original string was {\it supercritical}~\cite{aben}. 

The spatial part of the $\sigma$-model metric (\ref{ball})
may then be written in the form:
 \be
  G_{ij} = e^{-{\rm ln}(t^2 + |y_i|^2 )}
t^4 \delta_{ij} ~,i,j=1, \dots D~{\rm space-like}
\label{spacelike}
\ee
There is a unique way in which this metric can become a solution 
of standard Einstein's equation in a $D+1$ Universe,
with time (\ref{physical}) and time-like metric component
(\ref{temporal}). One should redefine the spatial part of the metric by:
\be
  G_{00}^{ph}=G_{00}~, \qquad 
G_{ij}^{ph} =e^{-\Phi (y_i, t)}G_{ij} = t^4 \delta_{ij},
\label{dilaton}
\ee
with
\be
\Phi (y_i, t) \equiv -{\rm ln}(t^2 + |y_i|^2 )
\label{dilaton2}
\ee
where $\Phi (y_i, t)$ may be regarded as a dilaton contribution. 
For the purposes of the present work, we assume that such a dilaton 
configuration is consistent with the world-sheet conformal invariance 
of the Liouville-dressed $\sigma$ model. 
At present an explicit check of this is beyond our control. 

The consistency of the 
resulting metric $G_{\mu\nu}^{ph}$, $\mu,\nu=1, \dots D+1$, 
with Einstein's equations has non-trivial consequences. 
Using (\ref{dilaton}), we seee that the physical 
Universe is of FRW type with a scale factor 
\be 
    R(t) = t^2 
\label{scale2}
\ee
This can be contrasted with the tree-level cosmological model
of~\cite{aben}, where a linear expansion was found as
$t \rightarrow \infty$. 
We see from (\ref{ccd}) and (\ref{physical}) 
that the Universe (\ref{dilaton}, \ref{scale2}) has a time-dependent 
vacuum energy $\Lambda(t)$ which relaxes to zero as:
\be 
      \Lambda (t) = \Lambda (0) /t^2 = 1/R(t)
\label{lambdaphys}
\ee
In accordance to the standard Einstein's equation,
this time-varying {\it positive} 
vacuum energy drives the cosmic expansion:
\be
\left(\frac{{\dot R}(t)}{R(t)} \right)^2 = \frac{1}{3} \Lambda (t) 
\label{einstein}
\ee
where the dot denotes a derivative with respect to the 
physical (Einstein) time $t$. 
{}From the point of view of the stringy $\sigma$ model, this result should 
be interpeted as meaning that the dilaton configuration and the rest of
stringy matter act together in such a way that the conformal invariance
conditions are satisfied, and the contribution of other
of the fields does not alter the low-energy Einstein dynamics
at late stages of the evolution of the Universe, the only remnant of
string matter being the time-dependent vacuum energy.

\section{Comparison with Observations}

In this section we 
compare the above result (\ref{scale2}),(\ref{lambdaphys}), 
with observational constraints on the cosmological constant.
As was already mentioned in the introduction, data
on large-scale structure formation~\cite{BF} favour the existence of
some form of vacuum energy, as well as conventional matter.
However, these data do not discriminate between a time-dependent
contribution to the vacuum energy, as derived in the previous
section, and a true cosmological constant. Some such
discrimination is provided by recent studies of high-redshift
supernovae~\cite{sn}. These measure the evolving geometry of the Universe
over most of its history, and hence constrain the cosmic equation of state 
from the era corresponding to redshift $z \simeq 1 $ to the present.
The question arises, therefore, whether these observations may
distinguish in principle or in practice between a true
cosmological constant and the variety of time-dependent vacuum
energy derived above within our
Liouville approach to $D$ brane recoil~\cite{dbrecoil,kanti}. 

We first review briefly the parametrization of~\cite{sn},
which is used in their analysis. 
The experimentally measured quantities are redshifts
$z$ defined as $\lambda/\lambda_0 \equiv 1/(1 + z) = R/R_0$ 
(where $\lambda$ denotes wavelength and $R$ scale factor, with the
subscript $0$ denoting quantities at the
present epoch), angular diameter distances $d_A = D/\theta$
(for astrophysical objects of proper sizes $D$ that are assumed to be
known),  
proper motion distances $d_M =u/{\dot \theta}$ (where $u$ is a transverse 
proper velocity and ${\dot \theta}$ an apparent angular motion), and 
luminosities $d_L$. 
There is a relation~\cite{turner} between
these observables that is model-independent:
\be
d_L=(1 +z)d_M =(1 +z)^2 d_A
\label{relation}
\ee
which allows one to make a fit with only two of these quantities,
conveniently chosen to be the redshift $z$ and the luminosity $d_L$. 
Using Einstein's equations in a FRW Universe, 
the luminosities can be related~\cite{turner} to the energy 
densities $\Omega _X$ for different material components $X$:
\be
d_L = \frac{c(1 + z)}{H_0\sqrt{\Omega _k}}{\rm sinn}\{\sqrt{\Omega_k}
\int _0^z dz' 
[\sum _{i} \Omega _i (1 + z')^{3(1 + \alpha _i)} + \Omega _k (1 + z')^2 ]^{-1/2}\}
\label{lumi}
\ee
where the $\Omega _i$ denote the normalized
energy densities of the various energy components, 
excluding the one corresponding to the spatial curvature,
and $\Omega _k = 1 - \sum _{i} \Omega _i$ 
denotes the effects of the 
spatial curvature of the FRW Universe.
The function ${\rm sinn}(x)$ is defined by
\bea 
{\rm sinn}(x) &=& {\rm sinh}(x)~ {\rm for} ~~ \Omega_k > 0 \nn \\
&=&   x ~ {\rm for} ~~ \Omega _k = 0 \nn \\
&=& {\rm sin}(x) ~ {\rm for } ~~ \Omega _k < 0 
\label{cases}
\eea
and the scaling exponents $\alpha _i$ are defined in terms 
of the pressure $P_i$. Specifically, 
for an energy component $\rho _X$ which scales like:
\be
     \rho _X \sim R^{-n} \qquad ; \qquad n=3(1 + \alpha _X) 
\label{cosmic}
\ee
where $R$ is the cosmic scale factor in a FRW Universe.
The analysis is based on an equation of state, derived from 
Einstein's equations, which defines $\alpha _X$ 
in terms of the pressure $P_X$:
\be
    \alpha _X  = P_X/\rho _X 
\label{state}
\ee
In the case of ordinary matter without a cosmological constant, 
$\alpha _{X=m}=0$, since the energy density of ordinary matter 
scales with the inverse of the spatial volume of the
Universe. On the other hand, in the case of a true
cosmological constant that does not vary with time, 
the constancy of the corresponding component of the energy density as the 
universe expands corresponds to $\alpha _{X=\Lambda} = -1$. 

The observational analysis of~\cite{sn}
constrained the cosmological equation of state of any unknown energy
component $\Omega_{X \ne m}$ that may contribute to 
the expansion of the Universe:
\bea 
 &~&   \alpha _X < -0.55 \qquad {\rm for~any~value~of~\Omega_m} \nn \\
&~&   \alpha _X < -0.60 \qquad {\rm for~~\Omega_m \ge 0.1} \nn \\
\label{ax}
\eea
The scaling of the
vacuum energy density given in (\ref{lambdaphys}),
which is inversely proportional to the scale factor, 
implies in the parametrization (\ref{cosmic}) of~\cite{sn}: 
\be
\alpha _\Lambda = -2/3
\label{ourmodel}
\ee
which is {\it consistent} with the observational high-redshift supernova
constraint (\ref{ax})~\footnote{We also note that
this is consistent with the null energy condition~\cite{wald},
which requires $\rho_X + P_X \ge 0$ and hence $\alpha_X \ge -1$.}. It is
encouraging that
the time dependence
we find is close to the range already excluded by the supernova
observations. This suggests that it may soon be possible to
exclude our speculative proposal.

A vacuum energy that relaxes to zero according to a general
power law
\be
\Lambda = \Lambda _0 /t^\lambda 
\label{relaxing}
\ee
is restricted by several phenomenological constraints.
Here we review some relevant considerations, with particular
emphasis on the specific features that are most relevant
to the recoil model described above.
We emphasize that our calculation is not a complete one, and
the contribution whose functional form we have discussed above
may not be the only contribution to the vacuum energy, and may
not even be the dominant one. However, for the purposes of this
discussion we assume that  the recoil contribution is indeed
dominant.

Being inspired by the superstring approach, 
which underlies our $D$-brane analysis,
we focus on theories which reduce to supergravity at large
distances. If supersymmetry were unbroken, the vacuum energy would be
zero, and one would expect a zero cosmological constant. 
However, in all physically relevant theories, supersymmetry is broken
in the observable sector, so a non-zero vacuum energy is to be
expected. In generic supergravity models, one has a maximum value 
\be
\Lambda \sim M_W^2 M_P^2 
\label{cosma}
\ee
where $M_W \sim 100~{\rm GeV}$ represents the electroweak scale.
We consider this maximal $\Lambda _0$ as a possible initial
value at small $t$ before the relaxation mechanism kicks in.
We further assume that supersymmetry breaking occurs at a characteristic
temperature  
\be
    T_a \sim \sqrt{M_WM_P}
\label{sba}
\ee
Alternatively, in certain no-scale models~\cite{noscale} one has 
that 
\be
\Lambda _0 \sim M_W^4 
\label{cosmb}
\ee
and the temperature at which supersymmetry breaking occurs is 
\be
 T_b \sim M_W 
\label{sbb}
\ee
Thus we consider two possible sets of initial conditions
for the relaxation of the vacuum energy: 
either (\ref{cosma}, \ref{sba}) or (\ref{cosmb}, \ref{sbb}).

The constraints coming from early cosmology
are easily satisfied if one assumes that the 
matter energy density
dominates over the vacuum energy density 
\be
G_N \rho _m \ge G_N \Lambda_0 /t^\lambda
\label{matcons}
\ee
We first analyse the constraint (\ref{matcons}) 
in case of generic supergravity models (\ref{cosma}, \ref{sba}). 
We assume that the matter energy
density scales with temperature as $T^4$ at early epochs, and hence that
$t \sim T^{-2}$ in the Einstein frame,
in natural units.   
{}From this and (\ref{matcons}, \ref{relaxing}) 
we find
\be
        T^{2-\lambda}M_P^\lambda \ge M_W M_P 
\label{scalea}
\ee
It is clear that if we had $\lambda = 1$ we would need
$T \ge M_W$ for the temperature in late Universe,
which is clearly unacceptable. Fortunately, this is not the
relaxation rate we found above, which was
$\lambda =2$. For this case, the inequality
(\ref{scalea}) is always respected.

In the case of no-scale models~\cite{noscale},
the constraint (\ref{matcons}) leads to 
\be
       T^{2-\lambda}M_P^\lambda  \ge M_W^2 
\label{scaleb}
\ee
The case $\lambda=1$ leads to $T \ge 0.1$~K,
whilst the case $\lambda =2$ again
always satisfies the constraint (\ref{scaleb}).

We conclude that our model of a relaxing vacuum energy is compatible with
all the relevant observational and phenomenological constraints.

\section{Conclusions}

We have presented in this paper a heuristic calculation of a
contribution to cosmological vacuum energy $\Lambda \sim 1/t^2$.
This calculation is incomplete and unsatisfactory in many respects.
For example, we are unable to control all other possible string- (or $M$-)
theory contributions to the vacuum energy, and hence cannot be sure that
the contribution we have identified here cannot be cancelled or modified
by some other effect. Even within our approach, the calculation
presented here may well be invalid because our dilute-gas
approximation is unjustified or inadequate. Nevertheless, we think
that our result has several interesting features.

It exemplifies the possibility that the vacuum energy may be neither zero
nor a non-zero constant, but may instead be relaxing towards an asymptotic
value. This calculation reflects the philosophy that the vacuum should be
regarded as a dynamical medium in constant interaction with the matter
propagating through it, which induces recoil effects that should not be
neglected. The energy of quantum space-time foam is increased by this
recoil excitation, which vanishes only when the Universe becomes empty at
large times. 

We leave to future work the tasks of justifying such a
calculation more formally, of searching for possible cancelling
contributions to the vacuum energy, of determining the possible
asymptotic value of the vacuum energy, of going beyond the
dilute-gas approximation, incorporating features of realistic
string- ($M$-) theory models such as supersymmetry, etc.. However,
we are not discouraged by the fact that this simple-minded
calculation produces a result that is not in obvious
contradiction with observational data. If nothing else, perhaps
our calculation will stimulate attempts to pin down more accurately
the equation of state of the vacuum, which may not be trivial.

\vfill
\noindent {\Large \bf Acknowledgements}

This work was supported in part by
a P.P.A.R.C. advanced fellowship (N.E.M.)
and D.O.E. Grant
DE-FG03-95-ER-40917 (D.V.N.).


\begin{thebibliography}{99} 

\bibitem{einstein} A. Einstein, {\it Relativity: The Special and General
Theory: A Popular Exposition}, translated by R.W. Lawson 
(Crown, New York, 1961). 



\bibitem{turner} S. Carroll, M. Turner and H. Press, 
Ann. Rev. Astron. Astrophys. 30 (1992), 499, and references therein. 



\bibitem{cosmo} For theoretical reviews, see: S. Weinberg, Rev. Mod. Phys.
61 (1989), 1; and astro-ph/9610044. \\
For a recent review on experimental 
and theoretical bounds on the cosmological constant, see: 
H. Martel, P. R. Shapiro and S. Weinberg, astro-ph/9701099.

\bibitem{relaxation} For a representative sample of references 
on scenarios with a vacuum energy that relaxes to zero, see; \\
M. \"Ozer and M.O. Taha, Phys. Lett. 171B (1986), 363; Nucl. Phys. 
B287 (1987), 776; Mod. Phys. Lett. A13 (1998), 571; \\
M. Reuter and C. Wetterich, Phys. Lett. B188 ( 1987) 38; \\
C. Wetterich, hep-th/9408025 and references therein; \\
J. Lopez and D.V. Nanopoulos, Mod. Phys. Lett. A9 (1994), 2755;
{\it ibid.} A11 (1996), 1; \\
I. Zlatev, L.-M. Wang and P.J. Steinhardt, astro-ph/9807002.

\bibitem{nosusy} E. Witten, Mod. Phys. Lett. A10 (1995), 2153. 

\bibitem{nosusystr} S. Kachru, J. Kumar and E. Silverstein,
hep-th/9807076; \\
S. Kachru and E. Silverstein, hep-th/9810129.

\bibitem{COBE} For a review, see: C. Lineweaver, astro-ph/9810334.

\bibitem{BF} For a review, see: N.A. Bahcall and X.-H. Fan,
astro-ph/9804082. 

\bibitem{GS} For a review, see: E. Gawiser and J. Silk, Science 280
(1998), 1405.

\bibitem{SK} Super-Kamiokande Collaboration, Y. Fukuda {\it et al.},
Phys. Rev. Lett. 81 (1998), 1562.

\bibitem{sn} S. Perlmutter {\it et al.}, astro-ph/9712212; \\
A.G. Riess {\it et al.}, astro-ph/9805201; \\
P. Garnavich {\it et al.}, astro-ph/9806396. 


\bibitem{dbrecoil} J. Ellis, N.E. Mavromatos and D.V. Nanopoulos, 
Mod. Phys. Lett. A12 (1997), 1759; 
Int. J. Mod. Phys. A12 (1997), 2639; 
{\it ibid.} A13 (1998), 1059.


\bibitem{kanti} J. Ellis, P. Kanti, N.E. Mavromatos, D.V. Nanopoulos
and E. Winstanley, Mod. Phys. A13 (1998), 303. 





\bibitem{dbranes} J. Polchinski, Phys. Rev. Lett. 75 (1995), 184; \\
C. Bachas, Phys. Lett. B374 (1996), 37; \\
J. Polchinski, S. Chaudhuri and C. Johnson, hep-th/9602052
and references therein; \\
J. Polchinski, TASI lectures on $D$ branes, hep-th/9611050, and references
therein; \\
E. Witten, Nucl. Phys. B460 (1996), 335. 





\bibitem{hilbert} D. Hilbert, Math. Phys. 53 (1917), 1. 

\bibitem{dewitt} B. DeWitt, in {\it Gravitation: An Introduction to 
Current Research}, edited by L. Witten (Wiley, New York, 1962); 
Phys. Rev. {160} (1967), 1113. 

\bibitem{rovelli} C. Rovelli, Class. Quant. Grav. {8} (1991), 297; 
{\it ibid. } 317;
\par J.D. Brown and D. Marolf, Phys. Rev. D53 (1996), 1835. 





\bibitem{mth} J. Ellis, N.E. Mavromatos and  D.V. Nanopoulos, 
hep-th/9804084, Int. J. Mod. Phys. A in press.

\bibitem{kmw} I. Kogan, N.E. Mavromatos and J.F. Wheater, Phys. Lett. B387 (1996), 483. 

\bibitem{gurarie}  V. Gurarie, Nucl. Phys. B410 (1993) 535;\\ M.A.I.
Flohr, Int. J. Mod. Phys. A11 (1996) 4147; {\it ibid.} A12 (1997) 1943;\\ M.R.
Gaberdiel and H.G. Kausch, Nucl. Phys. B489 (1996) 293; Phys. Lett. 
B386 (1996) 131;\\ F. Rohsiepe, hep-th/9611160;\\ I.I. Kogan, A.
Lewis and O.A. Soloviev, Int. J. Mod. Phys. A13 (1998) 1345. \\
For applications relevant to our context, see: \\
A. Bilal and I. Kogan, Nucl. Phys. B449 (1995), 569; \\
I. Kogan and N.E. Mavromatos, Phys. Lett. B375 (1996), 11; \\
J.S. Caux, I. Kogan and A.M. Tsvelik, Nucl. Phys. B466 (1996), 444;
\\  
N.E. Mavromatos and R.J. Szabo, Phys. Lett. B430 (1998), 94 and 
hep-th/9808124.


\bibitem{lizzi} F. Lizzi and N.E. Mavromatos, Phys. Rev. D55 (1997), 7859. 



\bibitem{ddk} F. David, Mod. Phys. Lett. A3 (1988), 1651; \\
J. Distler and H. Kawai, Nucl. Phys. B321 (1989), 509; \\
see also: N.E. Mavromatos and J. L. Miramontes, Mod. Phys. Lett. 
A4 (1989), 1847. 



\bibitem{emn} J. Ellis, N.E. Mavromatos
and D.V. Nanopoulos, Phys. Lett. B293 (1992), 37;
Mod. Phys. Lett. A10 (1995), 425; \\
Lectures presented at the
{\it Erice Summer School, 31st Course: From Supersymmetry to the
Origin of Space-Time},
Ettore Majorana Centre, Erice, July 4-12
1993 ; hep-th/9403133, `Subnuclear Series' Vol. 31, 
(World Scientific, Singapore 1994), p.1. 





\bibitem{emnmonop} J. Ellis, N.E. Mavromatos and D.V. Nanopoulos, 
Phys. Lett. B289 (1992), 25. 




\bibitem{fischler} W. Fischler and L. Susskind, Phys. Lett. 
B171 (1986), 383; {\it ibid.} B173 (1986), 262. 



\bibitem{aben} I. Antoniadis, C. Bachas, J. Ellis and D.V. Nanopoulos, 
Phys. Lett. B211 (1988), 383; Nucl. Phys. B328 (1989), 117. 




\bibitem{wald} R. M. Wald, {\it General Relativity} (University of Chicago 
Press, Chicago, 1984). 

\bibitem{noscale} A.B. Lahanas and D.V. Nanopoulos, Phys. Rep. 145 (1987), 1,
and references therein. 



\end{thebibliography}
\end{document}